\documentclass[12pt]{article}

\newcommand{\be}{\begin{equation}}
\newcommand{\ee}{\end{equation}}
\begin{document}
\title{A GENERAL RELATIVISTIC EFFECT IN  QUASI-SPHERICAL OBJECTS AS THE
POSSIBLE ORIGIN OF RELATIVISTIC JETS }
\author{L. Herrera$^1$\thanks{Postal
address: Apartado 80793, Caracas 1080A, Venezuela.} \thanks{e-mail:
laherrera@cantv.net.ve} and
N. O. Santos$^{2,3}$\thanks{e-mail: santos@ccr.jussieu.fr}
 \\
{\small $^1$Escuela de F\'{\i}sica, Facultad de Ciencias,}\\
{\small Universidad Central de Venezuela, Caracas, Venezuela.} \\
{\small $^2$LRM-CNRS/UMR 8540, Universit\'e Pierre et Marie Curie, ERGA,}\\
{\small  Bo\^\i te 142, 4 place Jussieu, 75005 Paris Cedex 05, France.}\\
{\small $^3$Laborat\'orio Nacional de Computa\c{c}\~ao Cient\'{\i}fica,}\\
{\small 25651-070 Petr\'opolis RJ, Brazil.}\\
}
\maketitle
\begin{abstract}
Based on recently reported results, we present arguments indicating that sign changes
in proper acceleration of test particles on the symmetry axis and close to the
$r=2M$ surface of quasi-spherical objects - related to the quadrupole moment
of the source - might be at the origin of relativistic jets of quasars and
micro-quasars.
\end{abstract}
\newpage

\section{Jets and quasi-sphericity}
Relativistic jets are highly energetic phenomena which have been observed
in many systems including active galaxies, X-ray binaries black holes, X-ray
transients, supersoft X-ray
sources, etc. (see \cite{Margon} and references therein), usually
associated with the
presence of a compact object. Despite the difference in typical scales of
jets for different cases, which range from light
years for the micro-quasars and million of light years for quasars, they
have analogous properties.

So far, the spin and the magnetic field of a compact object have appeared to be
the most popular candidates in powering and collimating jets
\cite{Blandford}. However, no consensus has
been reached until now, concerning the basic mechanism for its origin
\cite{Livio}. Furthermore, the great number of models available (see
\cite{Blandford} and references therein) and their complexity, implying
a large number of assumptions, indicate that the physics
underlying these jets have not quite been
understood and the question about its origin has not been answered yet.

In this letter we want to call the attention to a possible explanation to at
least some of these jets, which is based on the sign change of radial
proper acceleration of test
particles along the symmetry axis, close to the horizon of compact objects,
and  related to the quadrupole moment of the source \cite{Herrera} (although strictly speaking the term ``horizon'' refers to the spherically symmetric case, we
shall use it when considering the $r=2M$ surface, in the case of small deviations from sphericity).

Let us present our arguments. First of all, it should be reminded that in
the context of General
Relativity the only static and asymptotically flat vacuum spacetime
possessing a regular horizon is the
spherically symmetric Schwarzschild solution \cite{Israel}. If the field
is not particularly intense (the boundary of the source is
much larger than the horizon) and the deviation from spherical symmetry is
small, then it will be possible to
represent the corresponding fields, both inside and outside the source,
as a suitable perturbation of the spherically symmetric exact solution.
However, as the object becomes more
and more compact, such perturbative scheme will eventually fail near
the source. Indeed, as it is well known \cite{4}, as the boundary surface of
the source approaches the horizon,
any finite perturbation of the Schwarzschild spacetime, becomes
fundamentally different
from any Weyl metric, even if the latter is characterized by parameters
whose values are arbitrarily close to those corresponding to spherical
symmetry. In other words, for strong gravitational fields, no matter how
small the multipole moments (higher than monopole)
of the source are, there is a bifurcation between the perturbed Schwarzschild metric and all the other Weyl metrics.

Therefore if one wishes to describe the gravitational field of a
quasi-spherical source close to the horizon, one must use an
exact solution of Einstein equations,
rather than a perturbed Schwarzschild, no matter how small the
non-sphericity might be. If, for simplicity, one restricts oneself to the
family of axially symmetric non-rotating sources, then one has to choose
among the Weyl solutions.

However, since there are as many
different Weyl solutions as there are different harmonic functions, then
the  obvious question arises: what is the exact vacuum
solution of Einstein equations
corresponding to a given static axially symmetric source
(an ellipsoid, for instance)? Or more specifically: which one among the
Weyl solutions is
better entitled to describe small deviations from spherical symmetry?

These questions were partially answered in \cite{yo}, in which the exact
solution is presented and referred to as the $M$-$Q$ solution. And when
this solution has small values of $q=Q/M^3$, where $M$ stands for the mass
of the source and $Q$ its quadrupole moment, up to the first order of $q$,
it may be interpreted as the gravitational field outside a quasi-spherical
source.
This spacetime represents a quadrupole correction to the monopole
(Schwarzschild) solution. However this spacetime is in contrast with other
well known solutions
of the Weyl family, where
all the moments higher than the quadrupole present the same order of
magnitude in $q$. Therefore for small values of $q$ these solutions cannot
be
interpreted as a quadrupole perturbation of spherical symmetry.

In a recent study on the motion of test particles in the $M$-$Q$ spacetime
\cite{Herrera}, it was shown that particles moving along radial geodesics
experience an attractive force, as
one expects, except when they move along the symmetry axis. In this latter
case a repulsive force measured by a locally Minkowskian observer
 - regarded as a positive radial
proper acceleration measured by this observer - appears close to the
horizon if
$q<0$ (which corresponds to an oblate source) and even if it is arbitrarily
small
(though different from zero).
Indeed, first we define the local coordinates ($X$,$T$) associated with a
locally
Minkowskian observer or alternatively a tetrad field associated with such
Minkowskian observer,
\begin{equation}
dX=\sqrt{-g_{rr}}dr,
\label{x}
\end{equation}
and
\begin{equation}
dT=\sqrt{g_{tt}}dt,
\label{t}
\end{equation}
with $r$ denoting the usual spherical radial coordinate. Then if the test
particle moves along a radial geodesics outside the symmetry axis for
$R\equiv r/M\rightarrow 2$, up to the first order in $q$, one will obtain
the
following equation of motion
\begin{equation}
\frac{d^2 X}{dT^2}\approx -\frac{1}{2^{3/2}M(R-2)^{1/2}}
\left[1-\left(\frac{dX}{dT}\right)^2\right]+
\frac{{\cal O} (q)}{(R-2)^{1/2}}.
\label{limit3}
\end{equation}
Observe that when $q=0$, (3) becomes the exact
spherically symmetric situation. Thus for particles moving along radial
geodesics, excluding the symmetry axis,
small values of the quadrupole moment introduce small perturbations on the
trajectories, and the attractive nature of the gravitational force is
preserved at all  times.

However, for test particles moving along the symmetry axis, it follows that in the limit as 
$R\rightarrow 2$, the leading term in the equation of motion is
\begin{equation}
\frac{d^2 X}{dT^2}\approx -
\frac{5q}{2^{7/2}M(R-2)^{3/2}}\left[1-\left(\frac{dX}{dT}\right)^2\right]
\exp\left[-\frac{5q}{8(R-2)}\right],
\label{limit3}
\end{equation}
which is a positive quantity if $q<0$, for any value of $q$ different from
zero, diverging at the horizon. It should be emphasized that at $r=2M$ there
are no locally Minkowskian
observers and, as expected, the proper radial acceleration diverges there.
However, it should be clear  that the sign change takes place at  a  distance arbitrarily close to,  but still larger than $2M$, where such
Minkowskian observers do exist.

Although we have not a clear cut explanation for this bifurcation in the behaviour of the test particle, in the vicinity of the symmetry axis.  We conjecture that it might be related to
the well known directional behaviour of  singularity, inherent to most Weyl metrics (see for example \cite{susan} and references therein). 

Therefore, small deviations
from spherical symmetry close to the horizon with
$q<0$  produce a positive radial acceleration of particles moving along the
symmetry axis, as measured by a Lorentzian
observer. Thus, under these circumstances,  the observer would infer the
existence of a {\it repulsive} gravitational force acting on the particle.
This conclusion is valid in
all locally Minkowskian frames, including, of course, the instantaneous rest
frame where $dX/dT=0$. For example, if a test
particle initially at rest with respect to the
Minkowskian local frame is placed close to the horizon, on the axis of
symmetry as seen by such observer, it will tend to move towards increasing
$X$ values, away from
the source. Observe that in
the case of prolate compact objects, $q>0$, there are no repulsive forces,
but the attractive gravitational
force tends to zero as the particle approaches the horizon, this fact on
the other hand,
would enhance the effect of any other ejection mechanism near that region.

This repulsive force diverges as
the particle approaches the horizon (when $q<0$),
thereby providing an almost unlimited quantity of energy for outflowing
jets. On the other hand, the mere fact that such repulsive force acts
exclusively on particles along the
symmetry axis on the neighborhood of the horizon, provides at once the
explanation to the observed collimation of jets and, furthermore, it is
consistent with the conical geometry of
the jet, which were recently proposed to interpret observations of SS 433
using the
Chandra High Energy Transmission Grating Spectrometer \cite{Marshall}. 

\section{Conclusions}
Relativistic jets are characterized by:
\begin{itemize}
\item The presence of a compact object.
\item Very high energies involved.
\item A remarkable collimation (with possible conical geometry).
\end{itemize}
Concluding, in the case of a compact object, when its boundary surface is
close to its horizon and assuming only general relativistic effects, small
quadrupole corrections to spherical
symmetry may produce sign changes of proper acceleration of particles
moving  along the symmetry
axis in the neighborhood of the horizon . This would result in outgoing
jets, which are highly collimated and very energetic. This result is
independent on the presence, or absence, of
any other factors and mechanisms which
could enhance the ejection. Parenthetically, besides the magnetized
accretion disk
\cite{Livio,livio1}, powerful jets seem to require
an additional energy source.

Particularly interesting, in the context of the model here proposed, is the
recently observed fact
\cite{giovannini} that the velocity jet structure close to the central core
indicates a
strong decelaration. 
This effect is fully
consistent with our model, since the appearance of positive proper radial
accelerations takes place only very close to the horizon and change sign
as the particle moves away from the source.

It is worth noting that repulsive forces in the context of General
Relativity have been already shown. The best known example is probably
the case included in the
Reissner-Nordstr\"{o}m solution (see for example \cite{papapetrou}). Also,
these repulsive forces appear in connection with rotating and/or unbounded
sources
\cite{repulsiveI} or as spurious phenomena associated with coordinate
effects (see \cite{Mc} and references therein).
However, in our case, the repulsive nature of the force is exhibited by
means of a quantity  measured by locally Minkowskian observers, and
therefore it is not a coordinate effect.
Also, the $M$-$Q$ spacetime is asymptotically flat and, as stated before,
physically meaningful. Furthermore there
is a plausible source for such a metric, not violating energy conditions \cite{mqsource}.

However, it could be argued that as the object becomes a black hole (when $R \rightarrow 2$) then according to the non--hair theorem \cite{price} any quadrupole ``hair'' must be
radiated away, and therefore close to the horizon we should assume $q=0$. Let us elaborate on this important point.

As stated before, we are considering the $M$-$Q$ solution as the spacetime produced by quadrupole perturbations of the spherically--symmetric (Schwarzschild) metric. These perturbations
of course take place all along the evolution of the object. Thus, even if it is true that close to the horizon, any of these perturbations is radiated away, it is likewise true that
this is a continuous process. In other words as a quadrupole ``hair'' is radiated away a new quadrupole perturbation appears which will be later radiated and so on. In other words,
assuming that quadrupole ``hairs'' are radiated away at some {\it finite} time scale, then at that time scale there will be always a quadrupole fluctuation producing the effect
described here.

Finally, it is also worth mentioning that we have restricted to the non--rotating case, even though  we expect rotation to be present in all compact objects. However, rotation 
is not only not expected to hinder, but rather to enhance the above mentioned effect.  Thus for example, the influence of general relativistic effects on the dynamics of jets, 
in the context  of Kerr spacetime, has been presented (see \cite{mashhoon} and references therein).

\end{document}